# A Flexible Encoding Model for Non-Unique Note Alignments

Suhit Chiruthapudi[1], Adam Štefunko[2], Silvan Peter[1], Patricia Hu[1], Jan Hajič jr.[2], Carlos Eduardo Cancino-Chacón[1]

[1] Institute of Computational Perception, Johannes Kepler University, Austria
[2] Charles University, Faculty of Mathematics and Physics, Institute of Formal and Applied Linguistics, Czech Republic

suhit.chiruthapudi@jku.at

## Introduction

Symbolic music alignment refers to the matching of a sequence of notes in a symbolically represented performance to those in a symbolically represented score. Various automated methods exist that perform note-level alignments (Chen et al., 2014; Gingras & McAdams, 2011; Nakamura et al., 2017), each producing uniquely matched note pairs. These pairs can be encoded in several formats; including MEI (Music Encoding Initiative et al., 2025; Devaney et al., 2023), MELD (Weigl et al., 2019), CSV (Peter et al., 2023), Corresp (Nakamura et al., 2017) and Match (Foscarin et al., 2022) files.

However, the requirement of unique pairing fails to capture the intricacies of various musical contexts. In rehearsal (Morsi et al., 2025), score segments may be repeated, reordered, or partially practiced. In forms of music such as basso continuo (Štefunko et al., 2025; Chiruthapudi et al., 2025), lead sheets or piano reductions, incomplete or indicative scores are commonplace.

In this abstract, we discuss the edge cases for non-unique note matches and propose a simple extension to the Match file format[1] (Foscarin et al., 2022) that allows for a more robust and flexible encoding of such matches. With limited changes such as the introduction of virtual pointer notes and annotation segments, we capture a wide variety of encoding issues without breaking backwards compatibility or introducing another format.

## 1 Extensions to the Match File

We introduce two new virtual pointer notes - *virtualSnote* and *virtualPnote* - to handle non-unique alignments where score and performance notes do not correspond one-to-one. We also propose to extend the existing section line in the Match format to allow for semantically meaningful annotations of sections. A summary of these proposed modifications are listed in Table 1.

[1] This format was first created by Gerhard Widmer, Emilios Cambouropoulos and Simon Dixon in the early 2000s, and later updated by Foscarin et al., (2022).

| Value Name | Value Type | Description | Example |
|---|---|---|---|
| **Virtual Score Note:** virtualSnote(*Anchor,AttributesList*) | | | |
| Anchor | string | Score note reference | n1 |
| AttrList | [string] | Comma-separated list of attributes describing the type of connection | [rehearsal_repetition,left_hand] |
| **Virtual Performance Note:** virtualPnote(*NoteID*) | | | |
| NoteID | string | Performance note reference | n42 |
| **Section:** section(*id,...,StartInPerfTime,EndInPerfTime,SectionAttributesList]* | | | |
| id | string | Identifier for the section | s1 |
| StartInPerfTime | int | Start time of performed section (MIDI ticks) | 7158 |
| EndInPerfTime | int | End time of performed section (MIDI ticks) | 9224 |
| SectionAttrList | [string] | Comma-separated list of section attributes | [rehearsal_repetition] |
| **Omitted Section:** omittedSection(*id,StartInBeatsUnfolded,EndInBeatsUnfolded,StartInBeatsOriginal, EndInBeatsOriginal,SectionAttributesList*) | | | |
| id | string | Identifier for the section | os1 |
| StartInBeatsUnfolded | float | Start of omitted unfolded score section (in beats) | 300.0 |
| EndInBeatsUnfolded | float | End of omitted unfolded score section (in beats) | 350.50 |
| StartInBeatsOriginal | float | Start of omitted folded score section (in beats) | 100.0 |
| EndInBeatsOriginal | float | End of omitted folded score section (in beats) | 150.50 |
| SectionAttrList | [string] | Comma-separated list of section attributes | [omitted_volta_bracket] |

**Table 1.** Description and examples of the parameters of *virtualSnotes*, *virtualPnotes* and *section*.

### 1.1 Virtual Score Notes

Virtual score notes are used to make alignments between a score note - which has already been aligned at least once with a performance note - and another performance note. The construction of a virtual score note connection to a performance note is as follows:

**virtualSnote**(*Anchor,AttrList*)-**note**(*NoteID,MIDIpitch,Onset,Offset,Velocity, Channel,Track*).

Virtual score notes cover cases where one score note aligns with multiple performance notes, such as repeated rehearsal passages. It can also encode a cluster of performance notes that correspond to a single score note, like in the case of a melodic jazz improvisation over a given chord, or a trill where it is unclear how many exact notes are contained within it.

### 1.2 Virtual Performance Notes

Virtual performance notes are used in cases where one performed note can refer to multiple score notes in the score. An example of such a case is where sections such as bars or phrases in the score - which have the same musical content - repeat at a later point in the score again. A performance of this musical content can correspond to any of these occurrences in the score, and each correspondence can be argued as a valid connection in the context of music rehearsal. The construction of a virtual performance note connection to a score note is as follows:

**snote**(*Anchor,[NoteName,Mod],Octave,Measure:Beat,Offset,Dur,OnsetInBeats, OffsetInBeats,ScoreAttrList*)-**virtualPnote**(*NoteID*).

The *virtualPnote* does not contain a corresponding AttributeList parameter such as the one in the *virtualSnote*, as the *ScoreAttrList* would already include any relevant information regarding that specific connection.

### 1.3 Modifications to the Section Line in the Match File

The section lines in the original Match format are used to demarcate sections in the score that are bounded by musical repeat signs. As the score file itself contains only a single occurrence of all such notes within such a section, when the score is unfolded to linearly lay out the music piece in its entirety, such sections can link the unfolded score notes to their corresponding folded score notes. The extended section line allows marking sections repeated in performance for reasons beyond notated repeats, such as practice or improvisation. The construction of the section line is as follows:

**section**(***id***,*StartInBeatsUnfolded,EndInBeatsUnfolded,StartInBeatsOriginal, EndInBeatsOriginal,* ***StartInPerfTime***, ***EndInPerfTime***, ***SectionAttrList***).

In cases such as rehearsal, only a specific section of the score might be performed, thereby rendering all other score notes as deletions. These deletion lines would unnecessarily increase the size of the Match file. We therefore introduce the **omittedSection** line to concisely demarcate the start and end times of such unplayed sections:

**omittedSection(*id,StartInBeatsUnfolded,EndInBeatsUnfolded,StartInBeatsOriginal, EndInBeatsOriginal,SectionAttrList*).**

## 2 Use-Case 1: Piano Rehearsal Alignment

In piano rehearsal, contrary to a polished performance, players frequently stop, repeat passages, or insert partial corrections, thereby deviating from the linear structure of the notated score. Repeated passages in rehearsal cause multiple performance fragments to map to the same score location. Rehearsal symbolic alignment therefore requires a more flexible encoding that is capable of handling non-linear progression, partial matches, and non-unique correspondences. Figure 1 illustrates a typical case of rehearsal alignment containing practice repetitions as well as errors such as pitch errors, mistakenly played extra notes (insertions) and unplayed notes from the score (deletions).

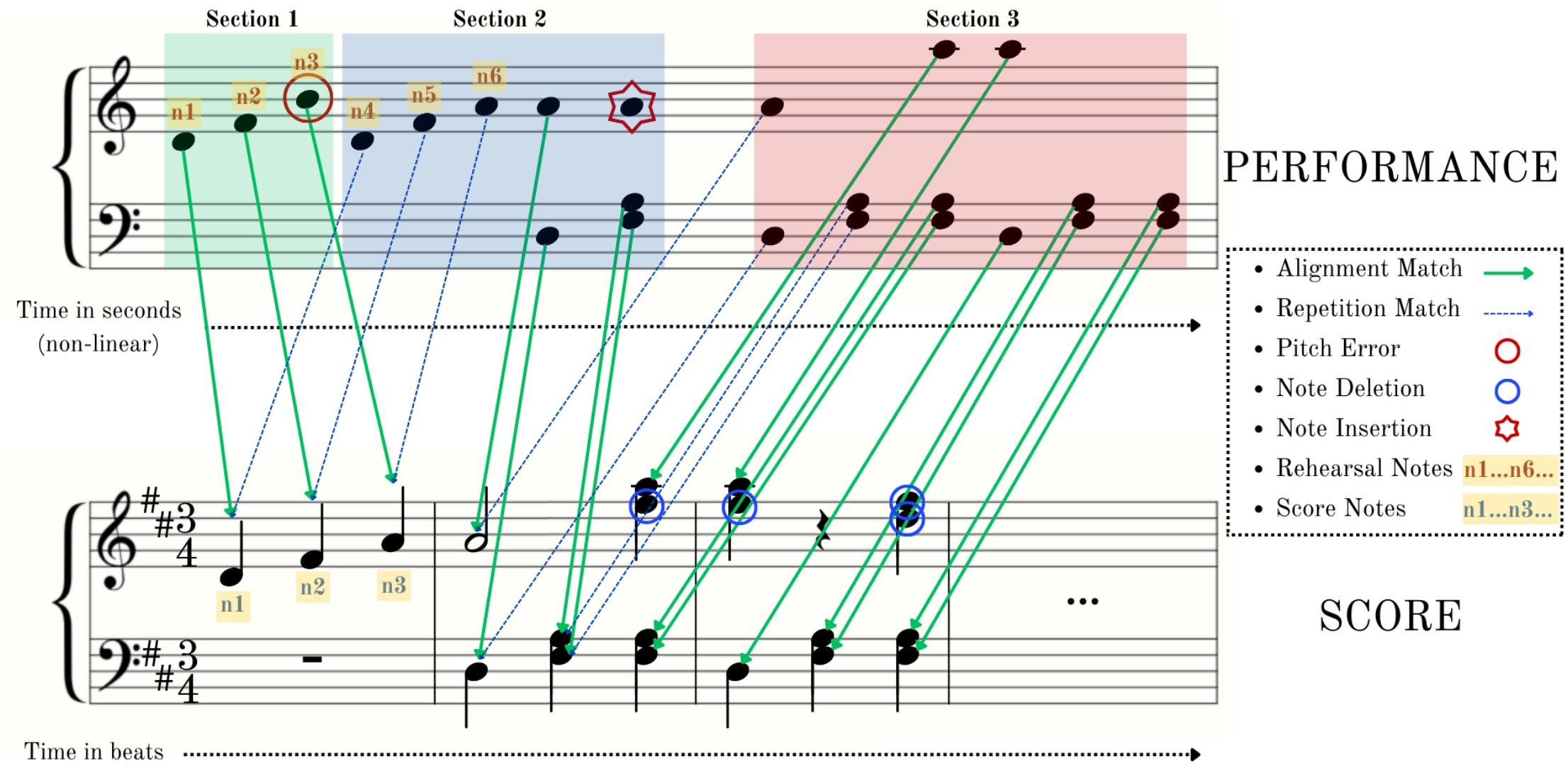

**Figure 1.** An example of the alignment of a piano rehearsal to its score. Sections of repeated practice corresponding to overlapping score regions are distinguished by colour, and arrowed lines align the performance notes to the score notes.

Figure 2 shows a fragment of the Match file that would be derived from the rehearsal alignment in Figure 1. It highlights the use-case of the proposed virtual score notes in encoding rehearsal repetitions of score passages, and the use of the modified section lines to semantically demarcate such sections in the rehearsal.

```
scoreprop(keySignature,D,1:1,0,0.0000).
scoreprop(timeSignature,3/4,1:1,0,0.0000).
section(s1,160.0,180.0,40.0,60.0,7586,9255,[rehearsal]).
snote(n1,[D,n],4,1:1,0,1/4,0.0000,1.0000,[v1,staff1])-note(n1,62,7586,8023,55,0,0).
snote(n2,[F,#],4,1:2,0,1/4,1.0000,2.0000,[v1,staff1])-note(n2,66,8191,8675,60,0,0).
snote(n3,[A,n],4,1:3,0,1/4,2.0000,3.0000,[v1,staff1])-note(n3,69,8798,9255,65,0,0).
section(s2,160.0,180.0,40.0,60.0,9387,11183,[rehearsal_repetition]).
virtualSnote(n1,["rehearsal repetition",c1])-note(n4,62,9387,9886,73,0,0).
virtualSnote(n2,["rehearsal repetition",c1])-note(n5,66,10012,10564,53,0,0).
virtualSnote(n3,["rehearsal repetition",c1])-note(n6,69,10712,11183,43,0,0).
```

**Figure 2.** The encoding derived from the rehearsal alignment in Figure 1. The lines corresponding to Sections 1 and 2 in Figure 1 are highlighted with the same colours (green and blue respectively). The corresponding score and performance Anchors and Note IDs in Figure 1 use the same colours for easy cross-reference.

Figure 3 illustrates an example where the rehearsal performance covers a bar of music that repeats twice in the score. It can be argued that the rehearsal of this bar covers the practice of both bar 1 and bar 9 in the score, and in order to allow for this flexibility, the virtual performance note can be used to link to both its corresponding score notes in bar 1 and 9. Figure 4 shows the corresponding Match file for this segment of rehearsal.

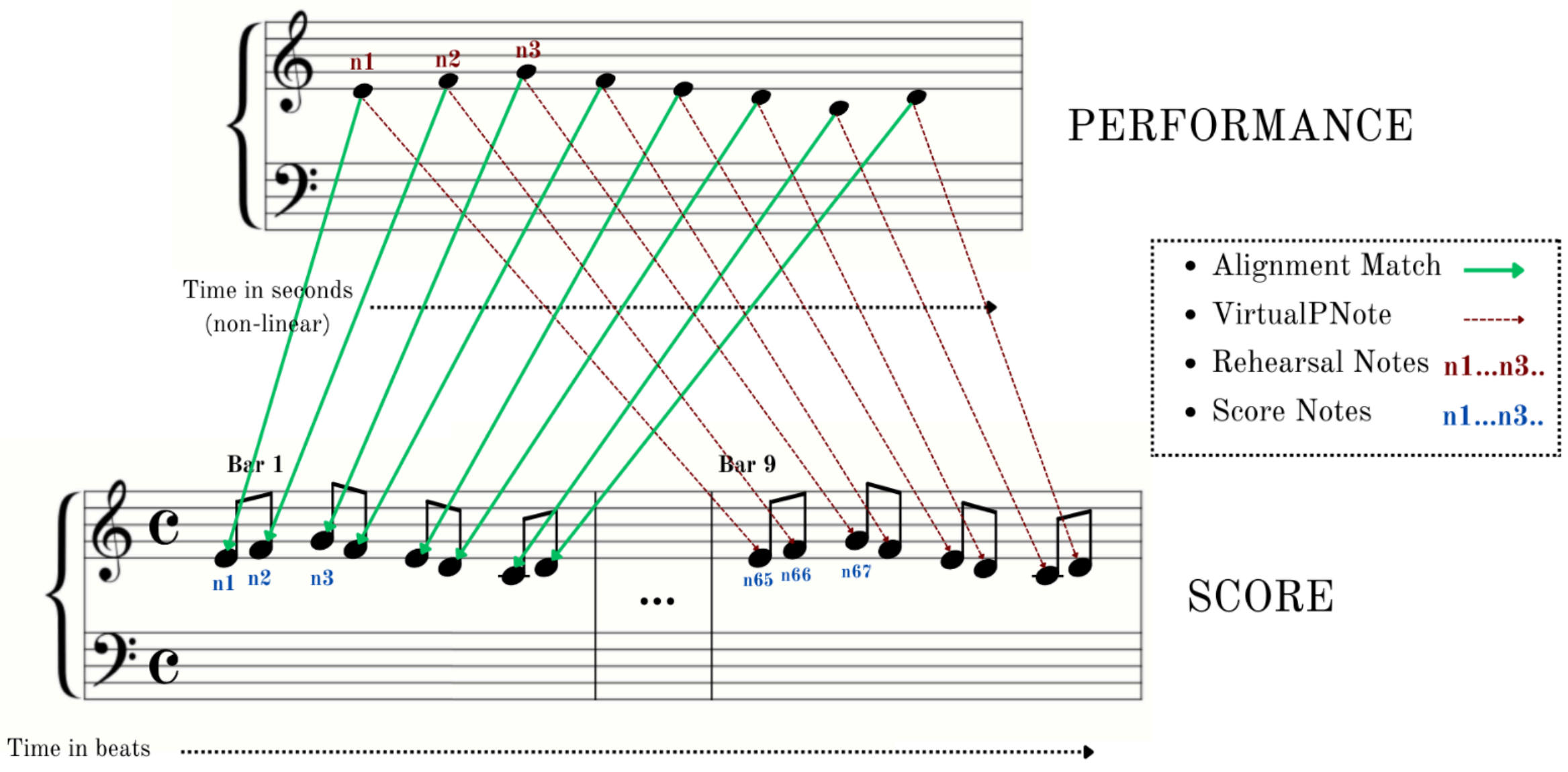

**Figure 3.** A sample example of the alignment of a segment of rehearsed music that corresponds to two different bars of music in the same score with the same musical content.

```
snote(n1,[E,n],4,1:1,0,1/8,0.0000,0.5000,[v1,staff1])-note(n1,64,1586,2023,55,0,0).
snote(n65,[E,n],4,9:1,0,1/8,32.0000,32.5000,[v1,staff1])-virtualPnote(n1).
snote(n2,[F,n],4,1:1,1/8,1/8,0.5000,1.0000,[v1,staff1])-note(n2,65,2191,2675,60,0,0).
snote(n66,[F,n],4,9:1,1/8,1/8,32.5000,33.0000,[v1,staff1])-virtualPnote(n2).
snote(n3,[G,n],4,1:2,0,1/8,1.0000,1.5000,[v1,staff1])-note(n3,67,2798,3255,65,0,0).
snote(n67,[G,n],4,9:2,0,1/8,33.0000,33.5000,[v1,staff1])-virtualPnote(n3).
```

**Figure 4.** The encoding of the alignment of performance notes n1, n2 & n3 in Figure 3 using virtual performance notes.

## 3 Use-Case 2: Basso Continuo Alignment

Basso continuo is the improvisatory keyboard accompaniment style of the baroque era, in which performers improvise parts above a written bass line, called *realization*. Aligning such performances to the notated bass requires mapping the performed bass note and multiple realization notes to a single score note, which would otherwise incorrectly appear as note insertions. Realization notes can also sustain across multiple bass notes, producing cases where one performance note needs to be aligned with multiple score notes. Both these cases are illustrated in Figure 5. In this context, virtual score and performance notes provide a semantically rich encoding of such alignments. Figure 6 shows the encoding of the notes enclosed in the dashed red box in Figure 5.

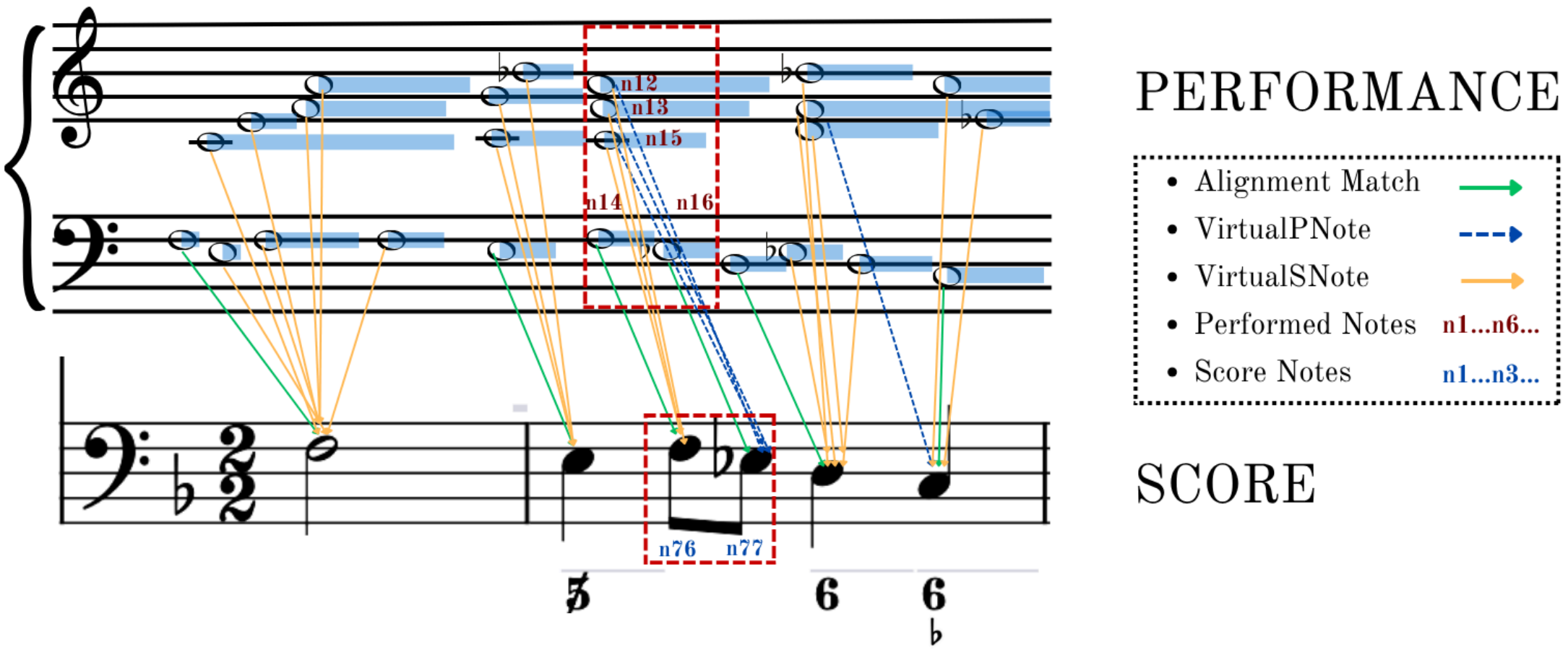

**Figure 5.** An example of a basso continuo performance-to-score alignment. The translucent blue bars extending horizontally from performance notes indicate their duration.

Although the virtual pointer notes in Figure 6 disrupt the performance-note order in the Match file ('n14' precedes 'n12'), it is worth noting that their flexibility allows the performed note sharing the score pitch to link to the original score note, preserving pitch matching.

```
snote(n76,[F,n],3,1:1,1/4,1/8,0.5000,0.750,[v1,staff2])-note(n14,53,2976,3262,1,0).
virtualSnote(n76,[realization])-note(n12,69,2964,3428,1,0).
virtualSnote(n76,[realization])-note(n13,65,2970,3396,1,0).
virtualSnote(n76,[realization])-note(n15,60,2984,3354,1,0).
snote(n77,[E,b],3,1:1,3/8,1/8,0.750,1.000,[v1,staff2])-note(n16,51,3221,3382,1,0).
virtualSnote(n77,[sustained_realization])-virtualPnote(n12).
virtualSnote(n77,[sustained_realization])-virtualPnote(n13).
virtualSnote(n77,[sustained_realization])-virtualPnote(n14).
virtualSnote(n77,[sustained_realization])-virtualPnote(n15).
```

**Figure 6.** The encoding of the notes in the enclosed region in Figure 5.

## Conclusion

We have outlined key alignment challenges that arise when score-performance relationships are non-unique, as in rehearsal practice and basso continuo performance. By introducing virtual score and performance notes and extending section annotations, we provide a simple, backward-compatible enhancement to the Match format that captures these complex correspondences.

## Acknowledgements

This work has been supported by the Austrian Science Fund (FWF), grant agreement PAT 8820923 (**Rach3: A Computational Approach to Study Piano Rehearsals**) and PIN 1347924 (**AURA: Augmenting musical interaction via EVAs**), as well as by the European Research Council (ERC) under the EU's Horizon 2020 research & innovation programme, grant agreement No.101019375 (**Whither Music?**).

This research was also partially supported by the project "**Human-centred AI for a Sustainable and Adaptive Society**" (reg. no.: CZ.02.01.01/00/23_025/0008691), co-funded by the European Union, by The Charles University Grant Agency (project no. 58126), and by the SVV project number 260~821. The computing infrastructure was provided by the LINDAT/CLARIAH-CZ Research Infrastructure (https://lindat.cz) supported by the Ministry of Education, Youth and Sports of the Czech Republic (Project No. LM2023062).

This research also acknowledges support by the European Research Council (ERC), under the European Union's Horizon 2020 research and innovation programme, grant agreement No. 101019375 Whither Music?.

## References

Chen, C.-T., Jang, J.-S. R., & Liou, W. (2014). Improved score-performance alignment algorithms on polyphonic music. In *2014 IEEE International Conference on Acoustics, Speech and Signal Processing (ICASSP)*, 1365–1369. https://doi.org/10.1109/icassp.2014.6853820

Gingras, B., & McAdams, S. (2011). Improved Score-performance Matching Using Both Structural and Temporal Information from MIDI Recordings. In *Journal of New Music Research*, *40*(1), 43–57. https://doi.org/10.1080/09298215.2010.545422

Nakamura, E., Yoshii, K., & Katayose, H. (2017). Performance Error Detection and Post-Processing for Fast and Accurate Symbolic Music Alignment. In *Proceedings of the 18th International Society for Music Information Retrieval Conference (ISMIR 2017)*. https://doi.org/10.5281/zenodo.1414940

Music Encoding Initiative, Roland, P., Kepper, J., Bohl, B. W., Cheng, J., Devaney, J., Dörner, S., Fujinaga, I., Hankinson, A., Kelnreiter, F., Komives, Z., Liska, U., Münnich, S., Pacha, A., Pfeffer, N., Plaksin, A., Porter, A., Pugin, L., Regimbal, J., ... Zhou, Y. (2025). music-encoding/music-encoding: MEI 5.1 (5.1). Zenodo. https://doi.org/10.5281/zenodo.14795947

Devaney, J., & Beauchamp, C. (2023). Encoding Performance Data in MEI with the Automatic Music Performance Analysis and Comparison Toolkit (AMPACT). arXiv:2311.11363

Foscarin, F., Karystinaios, E., Peter, S. D., Cancino-Chacon, C., Grachten, M., & Widmer, G. (2022). The Match File Format: Encoding Alignments Between Scores and Performances. In *Proceedings of the Music Encoding Conference 2022. Halifax, Canada.* https://doi.org/10.48550/ARXIV.2206.01104

Morsi, A., Chiruthapudi, S., Peter, S., Pilkov, I., Bishop, L., Maezawa, A., Serra, X., & Cancino-Chacón, C. (2025). Enabling empirical analysis of piano performance rehearsal with the Rach3 MIDI dataset. In *Proceedings of the 26th International Society for Music Information Retrieval Conference (ISMIR 2025)*, Daejeon, South Korea. https://doi.org/10.5281/zenodo.17811417

Štefunko, A., Chiruthapudi, S., Hajič, J. J., & Cancino-Chacón, C. E. (2025, September 10). Basso Continuo Goes Digital: Collecting and Aligning a Symbolic Dataset of Continuo Performance. In *The AI Music Creativity Conference (AIMC), Brussels, Belgium.* https://doi.org/10.5281/zenodo.16946799

Chiruthapudi, S., Štefunko, A., Hajič jr., J., & Cancino-Chacón, C. E. (2025). Challenges in Basso Continuo Performance-to-Score Alignment. In *Proceedings of the 17th International Symposium on Computer Music Multidisciplinary Research, 924–936.* https://doi.org/10.5281/zenodo.17502511

Weigl, D. M., Cancino-Chacón, C., Bonev, M., & Goebl W. (2019). Linking and visualising performance data and semantic music encodings in real-time. In *Late Breaking/Demo at the 20th International Society for Music Information Retrieval Conference, Delft, The Netherlands, 2019.* https://archives.ismir.net/ismir2019/latebreaking/000049.pdf

Peter, S., Cancino-Chacón, C. E., Foscarin, F., McLeod, A. P., Henkel, F., Karystinaios, E., & Widmer, G. (2023). Automatic Note-Level Score-to-Performance Alignments in the ASAP Dataset. In *Transactions of the International Society for Music Information Retrieval*, 6(1), *27 - 42.* https://doi.org/10.5334/TISMIR.149